# The role of gender in promotion rates in the Australian Finance Industry

# January 2023

Cassandra Crowe, CFA
cassandra.crowe@troweprice.com

Belinda Middleweek PhD
belinda.middleweek@uts.edu.au

Laura Ryan PhD (corresponding author)
laura.ryan@ardea.com.au

Alicia Vidler
a.vidler@unsw.edu.au

Bronwen Whiting PhD
bronwen.whiting@anu.edu.au

**Funding details:** The research project was conducted under a Gender Diversity Grant awarded by the Australian National University College of Business and Economics.

**Abstract**

We surveyed Australian finance professionals and tested whether there are statistically significant differences in promotional propensity according to gender identity. The findings indicate men and women are equally likely to ask for promotion, however, 'gifted advancements' account for the more statistically frequent promotion rates of men. These gender-based differences in behaviours have been overlooked in existing research on promotion. We call for a standardised framework for the development of promotion policies to address this industry-wide problem.

1. **Introduction**

Promotion is a re-evaluation of an employee's worth and serves, to use a finance term, to "mark to market", an employee on a regular basis. In financial markets, the process of "marking to market" captures the act of re-evaluating an asset's value at any given time. In this context we establish that promotions act as a way to update the employer's perceived value of an employee, relative to other employees and market dynamics. Promotions that carry with them a change in corporate title are arguably the most significant and most representative of individual merit. Promotions in the finance industry can occur in two ways: they can be initiated by the individual, or they can be offered to the individual. Organisations in finance have a number of mechanisms for evaluating an individual's suitability for a role which necessitate a change of title. For the purposes of this research, we have defined "promotion" as the strict definition of a change of title to a rank higher than currently held. Extant research has established women have a lower probability of promotion, and expected promotion despite controlling for performance and ability (Blau and Varo 2006) (Winchester, et al., 2006) (Ibarra, et al., 2010) (Behzad & Sharareh, 2021) (Murphy, et al., 2021). Regardless of how a promotion is awarded or applied, it is a metric that can also cut across segments of finance and specific roles to typify recognition of employee value and merit.

Research by CFA Institute shows women represent one in ten leadership positions globally across investment management in chief/c-suite roles (CFA Institute Research Foundation, 2016). This discrepancy does not go un-noticed with headlines such as "Fewer women run top Australian companies than men named John, or Peter or David"[1] being run across major Australian news outlets. Often the reasons put forward for this discrepancy include; a lack of female talent, (i.e. the pipeline argument) (Monroe & Chiu, 2010) (O'Brien, 2016) (Asare, 2018), women are too agreeable (Niederle. & Vesterlund, 2007) (Markman, 2012) (Risse, et al., 2018) (Risse, 2020) and career breaks due to family commitments (Garcia-Manglano, et al., 2014). These reasons however ignore the possibility of systemic discrimination and bias. It is for this reason that we have chosen to look at promotion through a gendered lens, that is, to test whether gender identity is a factor in promotion rates this study asks the question: are promotion rates independent of gender in the Australian finance industry?

---
[1] https://www.abc.net.au/news/2017-03-08/fewer-women-ceos-than-men-named-john/8327938

For the purposes of this study the Australian finance industry comprises employees across fund managers, financial advisers, superannuation funds and other institutions tasked with managing money and includes various technical and non-technical functions. The survey was comprehensive, (120 questions with an expected completion time of 20 minutes), guaranteed anonymity and asked detailed questions including historical and current salary details, career breaks, education, hours worked and promotion experience. The survey was circulated via email using the network of a large industry association (CFA Institute), the study's author's network within the Australian finance industry and LinkedIn to 2,000 employees in the Australian finance industry and 400 completed the survey in its entirety[2]. This was an unexpectedly positive outcome given the length of the survey and the personal nature of the questions being asked. 33.5% of whom identified as female. From the Australian government Workplace Gender Equality Agency known as WGEA approximately 4.2 million people in Australia are employed in the Finance industry, roughly 5% of the workforce in this industry are promoted annually, and 52.8% of employees are women (WGEA, 2021). The under-representation of women in the survey responses, compared to the sector, suggest that self-selection bias as a result of gender or gendered interest in the research was not a factor.

---

[2] At 90% confidence this translates to a required sample size of 271. The sample size for this study is 400. At 95% confidence this translates to 384. The results can therefore be assumed to be representative of the population. Please see the appendix for further discussion on model uncertainty, sample size and survey bias

The average survey respondent received AUD$284,357 in compensation in 2019. From (LU, 2022), average Australian income is $91,000; average figures for the finance and insurance industry are $121,332 (Male) and $96,345 (Female). Further, over 87% of respondents worked more than 40 hours per week normally. More than 96% of respondents were full time permanent employees, 49% of people held either a Masters or PhD and 78% had qualifications beyond a Bachelor's degree. It is clear their responses represent a significant portion of the highly skilled professionals engaged in managing money in Australia.

While promotion practices do exhibit differences across various financial firms, there are two basic pathways for achieving a promotion:

1. Be given a higher title following an application or request; or
2. Be given a higher title without having applied for or requested one (unsolicited).

With the survey respondents including 64.5% identifying as male, and 33.5% as female, 76% of the male finance industry professionals surveyed (who had ever received a promotion and who had not taken a career break) received an unsolicited promotion (pathway 2 above) compared with 60% of female respondents (see Table 1). When we further analysed these statistics for significance, we found promotion propensity was not independent of gender.

| Table 1: summation of the respondents, by gender, who had ever received any promotion | | | | |
|---|---|---|---|---|
| Gender | Ever Requested | Success rate of people when requested | Offered (without asking) | Ever Received |
| Female | 64.2% | 35% | 60.4% | 82.8% |
| Male | 68.6% | 21% | 76.0% | 90.3% |

We propose the term 'gifted advancement' to describe:

> **Gifted Advancement:** A promotion, dispensed in a beneficent manner, which is offered without prior request of the employee, at the decision of the employer, for the benefit of the recipient. This promotion carries with it a change of rank and increase of benefits to the employee.

Gifted advancement is a symptom of a bifurcated promotion process whereby male finance professionals, over the span of a career, benefit disproportionately from unsolicited promotions. The correlation between gender and promotion rates in the Australian finance industry is significant because the resulting compounded financial advantage is a direct result of less frequent promotion. We question the equitable nature of such a bifurcated system in the Australian finance industry and suggest future research explore the reasons for the promotion gap in the context of these findings.

The paper is laid out as follows, section 2 presents a literature review of the current work on gender promotion rates in finance and affiliated industries. Section 3 details the data. Section 4 explains the model and methods used to evaluate the results of survey data collected. In section 5 we present the results that augment existing research into the gender divide in promotion rates. We also detail the tale of two promotion types observed from data collected: the gifted promotion and the requested promotion. Section 6 contains a discussion of the impact of the results and section 7 concludes the paper.

2. **Literature review**

Much literature has been devoted to cataloging and detailing the difference in pay based on gender across numerous industries (Chamberlain, 2016) (CFA Institute Research Foundation, 2016) (Risse, et al., 2018). Significant work has established the quantifiable difference in pay in finance between men and women of the same occupation and industry (i.e. the adjusted pay gap), including in Australia, (Chamberlain, 2016) (Duong & Evans, 2016, pp. 17,32) (Levine, et al., 2017) (WGEA, 2022). focus specifically on the role of Chief Financial Officer (CFO) and analyse promotion and pay rates in Australia. They find female CFOs represent only 7.37% of firms in sample and that women CFOs are recognised for being "more conservative and deliver higher reporting quality compared to male CFOs". As a consequence, the authors argue female CFOs earn "significantly less than their male counterparts in all types of remuneration, even though they have obtained similar qualifications". Our work instead focuses on looking at a related, but distinct, feature of female professional attainment – promotion rates across the entire finance industry at all levels.

Looking at similar studies of similarly structured industries, (Azmat, et al., 2020, pp. 3, 1, 5) focused on law firm partnership levels in the US, with firms who used "transparent and homogeneous" measures for promotion to partner. They observed that "Twelve years after joining a law firm, women are 13% less likely to become partners than men". Noting that law school graduates and entrants to these firms were of an equal gender split, they suggested that women's promotion in transparent processes was affected by some personal attributes such as "aspiration". Crucially they found that female "aspirations are affected by early work experiences – facing harassment or demeaning comments early in the career affects long-term promotion outcomes mediated via aspirations". They went on to conclude that aspirations are "closely connected to the self-reported probability of becoming a partner". While some attribute the lack of women in leadership positions in U.S. law firms to individual aspirations of female employees, in U.S. retail settings the disadvantage is structural and linked to bias in the performance appraisal of female employees. (Benson, et al., 2021, pp. 0, 1, 26) looked at a large US retail firm where management tracked employees who went through a rigorous and transparent promotion process. They found "women receive substantially lower potential ratings "(rating of employee "potential") "despite receiving higher job performance ratings". They go on to conclude that "because 'potential' can never be directly observed, these assessments can be highly subjective, leaving room for bias". Furthermore, they observed that "even when women outperform their previously forecasted potential, their subsequent potential ratings remain low, suggesting that firms persistently underestimate the potential of their female employees" and "subjective assessments of worker potential contribute to gender gaps in promotion and to an inefficient allocation of talent across roles" with the final suggestion being that "despite being more likely to receive top performance ratings, women are less likely to be thought of as possessing high "potential". The utility of this finding is to provide a possible explanation for the results of our study.

Work such as (Bertrand, 2011) report that women are more risk averse, but "men appear particularly overconfident in their relative ability". Reports by (Barber & Odean, 2001) discuss male day traders' overconfidence compared to women. The facet of male "over confidence" in the workplace has relevance given the often-touted argument that women are not aggressive (confident) enough or too aggregable (lacking confidence) when it comes to requesting promotions. We utilise the seminal work of (Barber & Odean, 2001) to test whether women in the Australian finance industry are disinclined to ask for promotion because of a gendered personality trait. When looking at comparative work, according to (Acosta, 2004, p. 26), the idea that a formal promotion process is not based purely on merit is explored. Acosta argues that "those with a successful history at the firm do not seem to have an advantage when future promotions are made" and that it is possible but untested that "promotions are not based solely on productivity, but decided instead under other criteria or administrative rules different than performance, such as loyalty, influences, favouritism and other personal relationships, or by privileging the relationship between coworkers". When dealing explicitly with non-transparent or alternative promotion pathways we find two existing bodies of work exploring this. In their study of academic women's promotion in Australian Universities, (Winchester, et al., 2006, pp. 5-6, 505) found that "…whilst a strong official process of peer reviewed promotion existed" and "applications rates and success rates for women are similar to men's and, at professorial level, slightly higher". Significant difference was found in the informal process of "out of round promotions". This covered an "invisible" pathway to promotion open to those willing to engage in achieving a "counter-offer" from a competing university. As stated by one of the study participants such situations feature "no competition involved, there are no specifications as to what sort of offer is acceptable". The authors argue that "not one woman has been promoted out of round" whilst finding however that "it is quite a proportion of additional men".

The few studies that investigate unofficial promotions processes tend to conclude that unsolicited promotions reward male employees. For example, "men appear particularly overconfident in their relative ability" in the workplace (Bertrand, 2011), and so an unofficial promotions process without the checks and balances of agreed assessment criteria or other independent measure would necessarily advantage them over others. (Winchester, et al., 2006, p. 6) goes on to conclude that in this particular "invisible" process "women will not play the game"**.**

Aside from unofficial processes that privilege male employees, perceived competence also contributes to the lack of promotion of women in the workplace. (Moss-Racusina, et al., 2012) found both male and female evaluators rated the male candidate as more 'hireable' than the female candidate, even though the applications were identical outside of gender. In addition, higher starting salaries for the male applicant were recommended with additional mentoring opportunities, compounding this very real social and economic challenge.

Promotion rates are also impacted by dynamics occurring outside of the workplace. (Cullen & Perez-Truglia, n.d., pp. 0, 3) investigated the informal interactions of employees at a large commercial bank in Asia between 2015-2018 to estimate the "impact of social interactions on career progression". Specifically, they examined the prevalent cultural norm in the particular Asian region studied of nicotine smoking breaks as a means for managers and employees to interact in the workplace. The authors found a set of relationships between male managers and staff that demonstrated the stronger non-work time social smoking relationship positively affected promotion rates. Looking further they found that male manager to male staff member is an advantage explaining a third of the gender gap in promotions in this firm. But they note that male-to-male advantage is only present among employees "who work in close proximity to their managers". There also appear to be advantages of working for a manager who smokes if the employee also smokes. This is attributed to increased social interactions which "leads to significantly faster career progression" which is not accompanied by any differences in effort or performance. The authors note that this effect is not replicated with females, even those that smoke with a manager. The positive impact of social engagement is only felt by men. They reported that male employees were significantly more likely to share work breaks with a male manager after transferring from a female manager to a male manager.

The impact of bonds between males and their direct male managers is clearly established as a significant factor in explaining higher male promotion rates. Such a finding correlates with research that has found "employee potential" ratings favour male employees, even in the face of previous examples of female staff consistently exceeding their potential rating when reviewed at the end of the year (Benson, et al., 2021). It is these contrasting promotion rates that concern the current paper and is the driving force behind our request for the Australian finance industry to bring transparency and accountability to the promotion process.

As is the case in academic research, there are relatively few industry or organisational studies of the observed difference in rates of promotion between men and women. Further, to the author's knowledge there are no studies of promotion and pay procedures within the Australian finance industry. The authors are however employed within the finance industry and have been employed by various Australian and overseas based organisations. Typically we have observed non-uniform promotion and pay processes which include:

- Promotions determined by a committee whereby the manager of the employee argues on behalf of the employee for the pay rise or promotion. This is arguably the most common method observed and usually occurs once or twice a year. Promotions considered at these dates are usually referred to as "in round".
- Applications made by the employee to the manager who then either has the authority to grant a pay rise or promotion or can then decide whether to proceed with an application to the committee
- "out of round" promotions or pay rises usually occur where the employee has indicated that they have another offer or are considering leaving the firm.

Research by the CFA Institute shows women represent one in ten leadership positions globally across investment management in chief/c-suite roles (CEO, CIO, CFO)[3]. The Australian government agency WGEA publishes their own data showing women make up 30.8% of "Key management personnel" in the Australian financial industry data set. The notable dearth of women in senior leadership roles in investment management and large corporate institutions is well reported. Most recently Hesta, an Australian Super fund with a predominantly female membership base, reported in a 2021 published gender survey[4] of investment managers. They found that more women are working in investment management roles than ever before (up from 17% in 2018 to 22% in 2020), however the number remains low. Important to, though mostly overlooked, in organizational studies of promotion is whether gender is a factor in promotional propensity. By testing whether the propensity to request promotions differs between male and female employees in the Australian finance industry, we can better understand differences in career advancement. In the following section we outline the process for examining this propensity through a survey of Australian finance industry professionals.

### 3. Data

**3.1 Sample Size**

As of May 2021, 488,400 people were employed in the finance and insurance sectors[5]. According to the Australian Government's Labour Market Information Portal, we estimate the population relevant to our survey to be approximately 150,000. With a margin of error of 5% and 90% confidence, this translates to a required sample size of 271[6]. At 95% confidence this translates to a required sample size of 384. The sample size for this study is 400. We note that opinion polls in Australia are often in the range of 1,000 – 2,000 people as a representative sample for a voting population of approximately 17 million[7]. Similar sample sizes are also utilised for much larger populations such as the United States[8]. For their investment of time in the completion of a lengthy survey the authors wish to thank the respondents, whose identity is necessarily protected and private.

---

[3] CFA Institute Research Foundation, Gender Diversity Report (2016)
[4] https://www.hesta.com.au/about-us/media-centre/HESTA-gender-survey-investment-managers-2021.html (2011)
[5] https://lmip.gov.au/default.aspx?LMIP/GainInsights/IndustryInformation/FinancialandInsuranceServices
[6] https://www.ncbi.nlm.nih.gov/pmc/articles/PMC3409926/

### 3.2 Bias

Questions may be raised about sampling bias in respect of this data, and in particular nonresponse rates. (Davern, 2013) lists the following ways to deal with non-response bias:

1. Comparison of the sample and population
2. Follow-up analysis
3. Wave analysis
4. Passive/Active nonresponse analysis (focus groups)
5. Benchmarking
6. Replication

Access to the population is not achievable and the survey has not been repeated, as such methods 1, 2, 5 and 6 are not an option. The remaining approaches require identification of the survey respondent (the survey informing this analysis was anonymous). However, while it is impossible to eliminate biases, all attempts were made to mitigate documented issues with surveys and in particular non-response rates. Incentives have been identified as a tool for boosting non-response rates (Davern, 2013) (Singer & Cong, 2013) (Roycroft, et al., 2020). In an effort to boost response rates, respondents were offered an incentive for completion of the survey, being the provision of average salaries for various job types after the conclusion of the survey. Beyond this, the survey was titled "Gender, Merit and Career Progression", and participants were informed that the intended benefits of the study were to:

> "improve understanding of how gender and perceived merit interact with respect to career progression and employment in the finance industry. We hope this understanding will be used to ensure firms in the finance sector are continually benefiting from hiring and promoting the best candidates for each position."

---

[7] https://www.aec.gov.au/enrolling_to_vote/enrolment_stats/
[8] https://www.scientificamerican.com/article/howcan-a-poll-of-only-100/

Respondents to our survey were induced to reply by promise of receipt of summary data which could be accessed by respondents through a link provided only at the conclusion of the survey. Respondents were induced to reply to the survey honestly and fulsomely in the proposed interest of receiving such rare and personally valuable data. Such data release was timed to coincide with the time of year many financial market professionals have the opportunity to re-negotiate compensation for the year ahead. We believe this inducement was adequate to encourage participants with many different attitudes to the survey.

4. **Method and Model**

This study tests whether there were statistically significant differences between males and females with respect to asking and receiving promotions in the highly competitive and well remunerated finance industry in Australia. Specifically, survey respondents were asked three questions; How many times have you asked for a promotion?; How many times have you received a promotion? How many times have you been offered a promotion (without asking)? We tested statistically whether the answers to these questions were independent of gender identity[9]. We further analysed survey responses by years of experience[10], function and education to ensure the impact gender identity had on the propensity to request promotion were isolated from other possibly influential variables.

Our research tests four aspects of promotions for both genders. Whether the:

I. propensity to request promotions is independent of gender identity

II. propensity to be offered a promotion without asking is independent of gender identity

III. propensity to receive a promotion when applied for is independent of gender identity

IV. propensity to receive a promotion in any circumstance (whether requested or unrequested) is independent of gender identity

An example will help to illustrate. With respect to the propensity to ask for a promotion, the null and alternative hypothesis under investigation would be:

<u>*Null Hypothesis $H_0$:*</u>
*There is no relationship between gender & the propensity to request promotion*
*(they are independent)*

<u>*Alternative Hypothesis $H_a$:*</u>
*There is a relationship between gender & the propensity to request promotion*
*(they are not independent)*

A Chi-Square test is the usual method to test for independence between levels of two or more categorical variables. In this case the two categorical variables are gender identity and the frequency with which respondents asked for a promotion over the course of their career. The frequencies are grouped into the following categories: never, a few (1-5), many (6-10), a lot (10+) and more frequently than I can recall. The test compares the observed distribution of data against the distribution that would be observed if the distribution across levels of "asking for a promotion" is the same for both genders. The more the observed data distribution deviates from the expected distribution, the more support there is to reject the null hypothesis in favour of the alternative. The Chi-Square test does not require distributional assumptions, equality of variances across the categories and is invariant to heteroscedasticity. It also allows analysis across multiple groups (layers). In general, non-parametric tests such as the one we apply in this study should be used in the following instances:

1. Variables are categorical (nominal or ordinal).
2. Size of the groups are unequal
3. The original data were measured at an interval or ratio level, but violate one of the following assumptions of a parametric test:
   a. The distribution of the data was clearly not normal (very skewed or kurtotic)
   b. Assumptions of equal variance or homoscedasticity are not met
   c. The continuous data were collapsed into a small number of categories.
   (McHugh, 2013)

---

[9] Gender identity included: other, rather not specify, female or male

[10] Years of experience accounts for time out of the work force due to career breaks

The data under review in this study satisfy the conditions listed above.

To test whether the observed data fits the assumed model, a p-value is usually calculated based on the Chi-Square test statistics and compared to a selected "alpha" or significance level. We do not take this approach and discuss the limitations of the standard p-value and how we allow for these limitations in the appendix. It could be argued that some factors (such as years of experience, education and function) may influence the propensity to request a promotion. For example, as an employee progresses through their career it would be expected that their confidence increases and this may in turn lead to more requests for promotion. However there is also a large body of research which highlights the existence of age discrimination. See (Bora & Baumgartner, 2019) and (Donizzetti, 2019) for a review of the literature. Functional groups may also tend to exhibit their own idiosyncrasies as a function of size. For example, sales teams are often large in comparison to the more quantitative functions. Larger functions may offer more opportunity for promotion. To control for these variables, we added additional layers to the cross tabulation which included years of experience, education and function[11].

The same methodology for the question of whether gender is independent of the propensity to be offered a promotion with and without asking and the propensity to receive a promotion overall was applied.

5. **Results**

We test for four aspects of promotion by gender as listed in Section 4. Further, we look at the above statistics in two cohorts – those people (men and women) who report taking a career break (defined as time out of their job for any reason for over 6 months), and those who do not. There were 400 respondents in the survey in total. Of the 400, 162 people had taken a career break[12].

Many people will take career breaks in their lifetime for a number of reasons including time off, return to education and maternity leave or paternity leave. Key results from the Workplace Gender Equality Agency's 2019-20 gender equality scorecard (WGEA, 2020) show that over 80.8% of the finance industry they survey provide paid primary carers leave for both men and women. We asked survey participants if they had ever taken a period of career break, of duration 6 months or greater (for any reason whatsoever). Close to half of women and over

---

[11] Counts for some functions were too small to be significant
[12] A discussion as to the reason for the career break and the impact this may have on promotion propensity is beyond the scope of this paper but a possible area of future research

1/3rd of men reported taking time out of their career. This period of time was for any reason and not constrained to relate to any one reason such as maternity leave.

| Table 2: Summary of career breaks for all respondents | | |
|---|---|---|
| | No career break | Yes career break |
| Female | 26.72% | 45.28% |
| Male | 73.28% | 54.72% |

### 1. Summary of all tests for all respondents

Table 3 summarises the responses (by gender identity) to the three questions; How many times have you asked for a promotion?; How many times have you received a promotion? How many times have you been offered a promotion (without asking)?

| Table 3: Summary Statistics of Responses to the Three Survey Questions Regarding Promotions | | | | | | | | | | |
|---|---|---|---|---|---|---|---|---|---|---|
| | Female | | | | | Male | | | | |
| Question | Never | A few (0 - 5) | Many (5 -10) | A lot (10+) | More frequently than I recall | Never | A few (0 - 5) | Many (5 -10) | A lot (10+) | More frequently than I recall |
| How many times have you asked for a promotion (promotion means change in title)? | 35.82% | 61.94% | 0.74% | 1.49% | 0 | 31.51% | 61.48% | 5.84% | 1.17% | 0 |
| How many times have you received a promotion? | 17.16% | 76.12% | 6.72% | 0 | 0 | 9.73% | 75.49% | 12.84% | 1.56% | 0.39% |
| How many times have you been offered a promotion (without asking)? | 39.56% | 58.21% | 2.24% | 0 | 0 | 23.74% | 67.32% | 8.17% | 0.39% | 0.39% |

To provide the reader with another perspective of the results we have collapsed the categories into "Never" and "At least once", see Table 4.

| Table 4: Summary Statistics of Responses to the Three Survey Questions Regarding Promotions – Collapsed Categories | | | | |
|---|---|---|---|---|
| | Female | | Male | |
| Question | Never | At least once | Never | At least once |
| How many times have you asked for a promotion (promotion means change in title)? | 36% | 64% | 32% | 68% |
| How many times have you received a promotion? | 17% | 83% | 10% | 90% |
| How many times have you been offered a promotion (without asking)? | 40% | 60% | 24% | 76% |

Before undertaking any formal statistical tests, eyeballing the results leads us to suspect there is little difference between the genders with respect to their propensity to request promotions. But there does appear to be a difference in the overall number of promotions received and the propensity to be offered a promotion without asking – with men appearing to be more likely to receive promotions whether they asked for them or not.

2. **Statistical Summary for All Respondents (respondents who have and have not taken a career break)**

Here we present the statistical analysis for all respondents. We exclude the analysis for the other cohorts as the methodology is identical. Table 5 presents a summary of the responses to the question "How many times have you asked for a promotion?". The first question we want to answer is whether gender identity is independent of the propensity to request promotion.

| Table 5: Summary of counts for all respondents to request a promotion (proportions of each gender identity cohort in brackets) | | | | | |
|---|---|---|---|---|---|
| | Never | A few(1-5) | Many(5-10) | A lot (10+) | All |
| Female | 48 (35.82%) | 83 (61.94%) | 1 (0.75%) | 2 (1.49%) | 134 (100%) |
| Male | 81 (31.40%) | 159 (61.63%) | 15 (5.81%) | 3 (1.16%) | 258 (100%) |
| Rather not specify | 4 (50.00%) | 2 (25.00%) | 2 (25.00%) | 0 (0.00%) | 8 (100%) |
| All | 133 (33.25%) | 244 (61.00%) | 18 (4.50%) | 5 (1.25%) | 400 (100%) |

Given the explanation in section 4 of the chi square test, the null hypothesis for our purposes assumes the proportions in each promotion category, for each gender identity cohort should be equal. Or in other words, counts in each promotion category should be proportional to the total number of people in that category only. Put another way:

$$Prob(Number\ of\ times\ requested\ promotion = X)$$
$$=$$
$$Prob(Number\ of\ times\ requested\ promotion = X|Male)$$
$$=$$
$$Prob(Number\ of\ times\ requested\ promotion = X|Female)$$

**Hypothesis Test 1: Gender Identity and the propensity to request a promotion are independent among all respondents**

*Null Hypothesis $H_0$:*
*There is no relationship between gender & the propensity to request promotion*
*(they are independent)*

*Alternative Hypothesis $H_a$:*
*There is a relationship between gender & the propensity to request promotion*
*(they are dependent)*

To undertake the Chi-Square test we need to calculate the expected proportions in each cell of Table 5 above. We further collapsed the categories into: Never (0), a few (1-5), more than 5 to allow ease of analysis and to ensure there were enough counts in each category (Table 6). We proceed with these collapsed categories for the rest of the analysis across all groups. We also exclude the "Rather not specify" category and report proportions rather than counts so as to allow ease of comparison. The Chi-Squared test statistic is 4.22 with a bootstrapped p-value of 0.12. We decide we cannot reject the null hypothesis and conclude there is no significant difference in propensity for males and females to request promotions[13]. Note that given the discussion regarding the misuse and abuse of p-values we do not specify a degree of significance with which to compare the p-value to

| Table 6: Summary of proportions and expected proportions (in brackets) for all respondents with collapsed categories | | | |
|---|---|---|---|
| | Never | A few (1 - 5) | More than 5 |
| Female | 35.82% (33.28%) | 61.94% (60.97%) | 2.24% (5.75%) |
| Male | 31.39% (33.26%) | 61.3% (61.01%) | 6.98% (5.74%) |
| Rather not specify | 50.00% (33.75%) | 25% (61.25%) | 25.00% (6.25%) |

To further disentangle the impact gender has on promotion propensity from the impact of possibly influential factors such as experience and education, we also tested to see if these factors were equally distributed across gender. Applying a two-sample t-test, confidence interval and test for equal variances, we compared the distribution of experience by gender identity and found the two distributions to be equal. Applying a chi square test we find that the distribution of educational attainment is independent of gender. To control for function, years of experience and education within the chi square testing framework, we added additional layers to the cross tabulation[14]. In the interests of brevity, we present results for the Portfolio Management function only from herein. This function was also the most populated.

To ensure enough expected counts in each cell were high enough to enable the calculation of a test statistic, we further collapsed the propensity to request a promotion into "Never" and "At least once". Education was split into three categories, Sub Bachelor, Bachelor/Honors and Masters/Doctorate and years of experience was split into 0-10 years, 10-20 years and 20+ years. For example, Table 7 below shows the proportions and expected proportions for respondents in each cell for Portfolio Managers only, with 10-20 years of experience and who are in the Masters/Doctorate educational attainment group.

| Table 7: Proportions and expected proportions (in brackets) for all respondents in the portfolio management functional group, with 10-20 years of experience with a Masters/Doctorate with collapsed categories | | |
|---|---|---|
| | Never | At least once |
| Female | 20.00% (21.74%) | 80.00% (78.26%) |
| Male | 22.22% (21.74%) | 77.78% (78.26%) |

The bootstrapped p-value is 0.92 and so we decide we cannot reject the null hypothesis of no significant association between the variables. That is, after controlling for education, function and years of experience males and females request promotions at the same rate.

These formal statistical tests across the other functions confirm our "guestimates" and we found the propensity to request promotion to be independent of gender and the propensity to receive any promotion overall and the propensity to be offered a promotion without asking are related to gender. Men received more promotions than the same group of females. Furthermore, promotion propensity, when promotions are requested, is also dependent on gender. Women in the group are more likely to receive a promotion when they apply as compared to men in the same group. In contrast to existing literature regarding commercial firms, we find that the female cohort of our study request official promotion at the same rate as men (countering the "lack of aspiration" argument) and also receive them in higher rates than men in the case where women request a promotion.

---

[14] Counts for some functions were too small to be significant

Table 8 summarises the test results for the four promotion aspects listed previously across all respondents (i.e. both those who took a career break and those who didn't), those who did take a career break and those who didn't.

| Table 8: Summary of the test results for the four promotion aspects | | | | | | |
|---|---|---|---|---|---|---|
| Cohort | Everyone | | Respondents who did not take a career break | | Respondents who did take a career break | |
| Question and Outcome | Is promotion propensity independent of Gender? | Outcomes | Is promotion propensity independent of Gender? | Outcomes | Is promotion propensity independent of Gender? | Outcomes |
| **Propensity to request** (How many times have you asked for a promotion) | Yes. | Men and women ask for promotions in the same proportions | Yes | Men and women are asking for promotions in the same proportions. | No | Men who have taken a career break ask for promotions more than women who have taken a career break |
| **Propensity to be offered without asking** (How many times have you been offered a promotion without asking?) | No | Men receive more unsolicited promotions than women | No | Men receive more unsolicited promotions than women. | No | Men in this group are offered promotions without asking more than women in this group |
| **Propensity to receive a promotion when applied for** | No | Women receive more promotions when they request them | No | Women receive more promotions when they request them | No | Women receive more promotions when they request them |
| **Propensity to Receive** (How many times have you received a | No | Men receive more promotions than women overall regardless of whether the | No | Men receive more promotions than women | No | Men receive more promotions than women |

| promotion regardless of whether it was offered or requested?) | promotion was offered or requested. | | |
|---|---|---|---|

Looking at the cohort who have not taken a career break, the same picture emerges as for the group containing all respondents. Both genders are equally as likely to request promotions. However, men received more unsolicited promotions than women, resulting in more promotions in totality. Also, when women do apply for a formal promotion they have a higher chance of conversion than the men of the same group. This confounds the result that men still receive unsolicited promotions at a higher rate than the same women who have a higher chance of being awarded a promotion. More on this in the discussion section to follow. For brevity, from hereon we only present summaries of the analysis rather than the full data sets and statistical analysis.

When looking at the cohort who have taken a career break for any reason, we see a different set of results. Regardless of the reason for career break we see a different gender pattern emerge in the data. Gender plays a role in all promotion statistics. Women who have had a career break are less likely to ask for promotions than men who have had career breaks. This may be a case of the "motherhood penalty, fatherhood bonus" dichotomy (Budig, 2017) (Wei-hsin & Yuko, 2021).

## 6. Discussion

Our results show men and women have the same propensity to request promotion across all groups of respondents excluding those that took a career break. Given request rates are unrelated to gender (minus the cohort who have taken a career break) then we can interpret this to mean that women are in fact leaning into asking for promotions. These findings contradict the notion that the pay and promotion gap is in part due to differing levels of "agreeableness" between women and men, where "agreeableness" is usually construed to mean women do not request or put themselves forward (being deemed too concerned with appearing to be agreeable).

While the results indicate women and men request promotions at a similar rate, the same cannot be said for receiving promotions. We found women were slightly more likely to receive a promotion when they request it. Overall, however, women were less likely to receive a promotion (regardless of whether they requested it or not) when compared to men. Further, we found that men were receiving promotions without asking for them (unsolicited promotions) more often than women – accounting for the difference. Unsolicited promotion is, by definition, a professional situation where people receive greater responsibility and pay without request.

The data also allowed us to infer the success of men and women when they request promotions. The results tell us that there is a group of men and women who have received a promotion but had never been offered them in an unsolicited manner. Some interesting differences between the genders emerged when we further analysed the probability of success given a request for promotion, amongst this "request only" cohort. We found that men in the cohort of people who only ever asked for promotions had a lower probability of being promoted when they requested one (17.2%) as compared to women (25.9%). This means the propensity for men and women to be deemed worthy of promotion when they requested them were different between the genders.

Looking at the distribution of women overall receiving promotions, they are significantly underrepresented in the pool of people receiving unsolicited promotion. Given that around 95% of senior members of ASX 200 listed companies are men, it is possible to assume that managers conferring promotions are almost certainly likely to be male, in most, if not all, situations. The probability of men in the cohort being offered an unsolicited promotion was more than 25% higher than that of a women respondent suggesting there may be some degree of unconscious bias (Bourne, 2019). (ELSEVIER, n.d.) provide the following explanation of unconscious bias: "Unconscious bias (UB) arises from a feature of the human brain that helps us make decisions faster via a series of shortcuts. It shapes our perception of the world and our fellow human beings and can lead to us make questionable decisions. It means that we often end up treating people and situations based on unconscious generalizations and preconceptions rather than using a set of objective qualitative or quantitative parameters."

## 7. Conclusion

We conducted a survey of Australian finance employees with the goal of understanding whether gender was a significant factor in determining promotion rates. Survey respondents were asked three questions; How many times have you asked for a promotion?; How many times have you received a promotion? How many times have you been offered a promotion (without asking)? Allowing for model uncertainty and the problems associated with comparing p-values to arbitrary significance we utilised a monte carlo bootstrap technique to test whether the answers to these questions were independent of gender identity. After controlling for education, years of experience, career breaks and function we found that among the respondents, men receive promotions more often than women, women and men are as equally likely to request a promotion and men are more likely to be offered a promotion without asking. This was true across the respondents who had not taken a career break and the entire data set, i.e. including those who had a career break.

The survey was circulated with the assistance of the Australian branch of an international organisation (CFA Institute) and included participants with work history internationally. Due to the dominance of large multinational firms in the finance industry, there are very likely aspects of work culture in these organisations which are transnational, but the impact of local legislation and culture cannot be ignored. We do not have sufficient data to comment on whether the results would apply beyond the Australian context, and further research is required to test if these results generalise to other English-speaking countries, and then other countries more broadly.

By examining whether promotion propensities are equal across gender identity with a particular focus on the Australian finance industry, this study offers an original contribution to existing literature on workplace promotion. Our results directly challenge the notion that a major factor determining the pay and promotion gap in the Australian finance industry is the disparate level of "agreeableness" between men and women. Rather, the difference in promotion rates in the Australian finance industry can be attributed in part to "gifted advancement" a term we define as: "a promotion, dispensed in a beneficent manner, which is offered without prior request of the employee, at the decision of the employer, for the benefit of the recipient. This promotion carries with it a change of rank and increase of benefits to the employee." We therefore call on the Australian finance industry to develop and adopt a standardised framework for the development of corporate promotion policies with the specific goal of mitigating the systemic bias present in current promotion rates. An open and transparent framework for all promotion opportunities reduces inherent bias across promotion opportunities, allowing individuals to navigate their own career, applying for roles that match their experience, skill set and ambition.

## 8. Declaration of interests

None

## 9. References

Acosta, P., 2004. Promotions, State Dependence and Intrafirm Job Mobility: Evidence From Personnel Records. *North American Summer Meetings, Econometric Society,* Volume 585.

Anderson, D., Burnham, K., Gould, W. & Cherry, C., 2001. Concerns about finding effects that are actually spurious. *Wildlife Society Bulletin,* 29(1), pp. 311-316.

Andeson, D. & Burnham, K., 2003. *Model Selection and Multi-Model Inference.* s.l.:Springer.

Asare, J. G., 2018. *Five Reasons Why The Pipeline Problem Is Just A Myth.* [Online]
Available at: https://www.forbes.com/sites/janicegassam/2018/12/18/5-reasons-why-the-pipeline-problem-is-just-a-myth/?sh=1c862883227a

Austin, P., 2008. Using the Bootstrap to improve estimation and confidence intervals for regression coefficients selected using backwards variable elimination. *Statistics in Medicine,* Volume 27, pp. 3286-3300.

Austin, P., 2008. Using the Bootstrap to improve estimation and confidence intervals for regression coefficients selected using backwards variable elimination. *Statistics in Medicine,* Volume 27, pp. 3286-3300.

Azmat, G., Cuñat, V. & Henry, E., 2020. Gender Promotion Gaps: Career Aspirations and Workplace Discrimination. *Econometrics: Single Equation Models eJournal,* p. 3.

Barber, B. & Odean, T., 2001. Boys Will Be Boys: Gender, Overconfidence, And Common Stock Investment. *The Quarterly Journal of Economics,* 116(1), pp. 261-292.

Behzad, R. & Sharareh, K., 2021. A gender-based analysis of workforce promotion factors in U.S. transportation agencies. *Transportation Research Interdisciplinary Perspectives.*

Benson, A., Li, D. & Shue, K., 2021. "Potential" and the Gender Promotion Gap" (Working paper, MIT).

Bernanke, B., Gertler, M. & Watson, M., 2004. Reply: Oil Shocks and Aggregate Macroeconomic Behviour: The Role of Monetary Policy. *Money, Credit and Banking,* 36(2), pp. 287-291.

Bertrand, M., 2011. New Perspectives on Gender. *Handbook of Labor Economics,* Volume 4, pp. 1543-1590.

Bora, J. & Baumgartner, L., 2019. *Ageism in the Workplace: A Review of the Literature.* s.l., New Prairie Press.

Bourne, J., 2019. Unravelling the Concept of Unconscious Bias. *Race & Class,* 60(4), pp. 70-75.

Buckland, S. T., Burnham, K. P. & Augustin, N. H., 1997. Model selection: an integral part of inference. *Money, Credit and Ban,* 53(2), pp. 603-618.

Budig, M., 2017. *THE FATHERHOOD BONUS AND THE MOTHERHOOD PENALTY* [Interview] 2017.

CFA Institute Research Foundation, 2016. *Gender diversity in investment management: New research for practitioners on how to close the gender gap.* [Online]
Available at: https://www.cfainstitute.org/en/research/survey-reports/gender-diversity-report

Chamberlain, A., 2016. *Demystifying the Gender Pay Gap,* s.l.: Glassdoor.

Chatfield, C., 1995. Model Uncertainty, Data Mining and Statistical Inference. *the Royal Statistical Society,* pp. 419-466.

Chatfield, C., 1996. Model Uncertainty and Forecast Accuracy. *Journal of Forecasting,* 15(7), pp. 495-508.

Clyde, M., 2000. Model uncertainty and health effect studies for particulate matter. *Environmetrics,* 11(6), pp. 745-763.

Clyde, M. & George, E., 2004. Model Uncertainty. *Statistical Science,* 19(1), pp. 81-94.

Cullen, Z. & Perez-Truglia, R., n.d. The Old Boys' Club: Schmoozing and the Gender Gap. *SSRN Electronic Journal.*

Davern, M., 2013. Nonresponse Rates are a Problematic Indicator of Nonresponse Bias in Survey Research. *Health Services Research.*

Donizzetti, A., 2019. Ageism in an Aging Society: The Role of Knowledge, Anxiety about Aging, and Stereotypes in Yound Pople and Adults.. *Int J Environ Res Public Health,* 16(8).

Duong, L. & Evans, J., 2016. Gender differences in compensation and earning management: Evidence from Australian CFO. *Pacific-Basin Finance Journal,* Volume 40, p. 17.

Eicher, T., Papageorgiou, C. & Raftery, A., 2011. Default priors and predictive performance in Bayesian model averaging, with application to growth determinants. *Applied Econometrics,* 26(1), pp. 30-55.

ELSEVIER, n.d. *Unconscious bias.* [Online]
Available at: https://www.elsevier.com/en-au/open-science/science-and-society/unconscious-bias
[Accessed 10 August 2022].

Garcia-Manglano, J., Bianchi, S. & Kahn, J., 2014. Kahn JR, García-Manglano J, Bianchi SM. The Motherhood Penalty at Midlife: Long-Term Effects of Children on Women's Careers. *Journal of marriage and the family,* 76(1), pp. 56-72.

Ibarra, H., Carter, N. M. & Silva, C., 2010. *Why men still get more promotions than women,* s.l.: s.n.

LavouÃ, J. & Droz, P., 2009. Multimodel inference and Multimodel Averaging in Empirical Modeling of Occupational Exposure Levels. *Annals of Occupational Hygiene,* 53(2), pp. 173-180.

Lazar, N. A., Schirm, A. L. & Wasserstein, R. L., 2019. *Moving to a World Beyond "p<0.05".* s.l.:Taylor & Francis.

Levine, B., Moldavskaya, D., Xiong, K. & Doherty, J., 2017. *Global Gender Pay Equity An Examination of Gaps Outside the US.* [Online]
Available at: https://info.mercer.com/WWT-Full-Report.html

Lopez de Prado, M., 2018. The 10 Reasons Most Machine Learning Funds Fail. *Journalof Portfolio Management, Forthcoming, available at: https://ssrn.com/abstract=3104816.*

LU, G., 2022. *bosshunting.* [Online]
Available at: https://www.bosshunting.com.au/hustle/average-australian-salary/,

Markman, A., 2012. *Harvard Business Review.* [Online]
Available at: https://hbr.org/2012/02/are-successful-people-nice

McHugh, M., 2013. the chi-square test of independence. Biochem Med (Zagreb).. 23(2), pp. 143-9.

Miller, A., 1990. *Subset selection in regression.* London: Chapman and Hall.

Monroe, K. R. & Chiu, W. F., 2010. Gender Equality in the Academy: The Pipeline Problem. *Political Science and Politics,* 43(2), pp. 303-08.

Moss-Racusina, C. A., Dovidio, J. F. & Brescoll, V., 2012. Science Faculty's Subtle Gender Biases Favor Male Students. *." Proceedings of the National Academy of Sciences of the United States of America (PNAS),* 109(41), pp. 16474-16479.

Murphy, M., Callander, J. K., Dohan, D. & Grandis, J. R., 2021. Women's Experiences of Promotion and Tenure in Academic Medicine and Potential Implications for Gender Disparities in Career Advancement: A Qualitative Analysis. *JAMA Network Open,* pp. e2125843-e2125843.

Niederle. , M. & Vesterlund, L., 2007. "Do Women Shy Away from Competition? Do Men Compete Too Much?". *Quarterly Journal of Economics,* 122(3), pp. 1067-1101.

O'Brien, S. A., 2016. *What pipeline problem? Carnegie Mellon nears gender parity.* [Online]
Available at: https://money.cnn.com/2016/09/16/technology/carnegie-mellon-computer-science/index.html

Risse, L., 2020. Leaning in - Is higher confidence the key to women's career advancement?. *Australian Journal of Labour Economics,* pp. 43-77.

Risse, L., 2020. Leaning in: Is higher confidence the key to women's career advancement. *Australian Journal of Labour Economics,* 23(1), pp. 43-78.

Risse, L., Fry, T. & Farrell, L., 2018. Personality and pay: do gender gaps in confidence explain gender gaps in wages?. *Oxford Economic Papers,* 70(4), pp. 919-949.

Roycroft, J., Stanley, M. & Amaya, A., 2020. The Effectiveness of Incentives on Completion Rates, Data Quality, and Nonresponse Bias in a Probability-based Internet Panel Survey. *Field Methods,* 32(2), pp. 159-179.

Singer, E. & Cong, Y., 2013. The Use and Effects of Incentives in Surveys. *American Academy of Political and Social Science,* Volume 645, pp. 112-41.

Wei-hsin, Y. & Yuko, H., 2021. Motherhood Penalties and Fatherhood Premiums: Effects of Parenthood on Earnings Growth Within and Across Firms. *Demography,* 58(1), p. 247–272..

WGEA, 2020. *Key results from the Workplace Gender,* s.l.: s.n.

WGEA, 2021. [Online]
Available at: https://data.wgea.gov.au/?_ga=2.92499806.232388846.1660093938-1471838448.1658960361
[Accessed 2022].

WGEA, 2022. *Australia's gender equality scorecard,* s.l.: s.n.

Winchester, H., Shard, L. & Browning, L., 2006. Academic Women's Promotion in Australian Universities. *Employee Relations,* Volume 28, pp. 505-522.

Young, C., 2018. Model Uncertainty and the Crisis in Science. *Socius,* January.

Zucchini, W., 2000. An introduction to model selection. *Mathematical Psychology,* pp. 41-61.

# 10. Appendix
## a. Model uncertainty

In 2019, the American Statistical Association (Lazar, et al., 2019) wrote:

"If you're just arriving to the debate, here's a sampling of what not to do:

- Don't base your conclusions solely on whether an association or effect was found to be "statistically significant" (i.e., the p-value passed some arbitrary threshold such as $p < 0.05$).
- Don't believe that an association or effect exists just because it was statistically significant.
- Don't believe that an association or effect is absent just because it was not statistically significant.
- Don't believe that your p-value gives the probability that chance alone produced the observed association or effect or the probability that your test hypothesis is true.
- Don't conclude anything about scientific or practical importance based on statistical significance (or lack thereof)."

The issues above result due to the researcher's attempt to generalise a single sample to the population. Attempting to model the population comes with various sources of model uncertainty which include uncertainty about the functional form of the model and distributional assumptions. Model uncertainty is poorly recognised within the finance literature (Lopez de Prado, 2018). Some of the methods available to researchers to combat model uncertainty include: bootstrap estimation of coefficients and confidence intervals (Austin, 2008), bootstrap model averaging (Buckland, et al., 1997), bayesian model averaging, (Clyde, 2000), and multi-model inference (Andeson & Burnham, 2003). (Eicher, et al., 2011) demonstrate improved predictive performance and (Lavouà & Droz, 2009) use Bootstrap and Bayesian model averaging (multi model inferencing). Given the discussion above, we also utilise a monte carlo bootstrap technique for determining the p-value for our hypothesis tests. We also utilised confidence intervals as a further attempt to allow for the model uncertainty issues and problems associated with comparing p-values to arbitrary significance levels as highlighted by the American statistical association.

**Tables**

| Table 1: summation of the respondents, by gender, who had ever received any promotion | | | | |
|---|---|---|---|---|
| Gender | Ever Requested | Success rate of people when requested | Offered | Ever Received |
| Female | 64.2% | 35% | 60.4% | 82.8% |
| Male | 68.6% | 21% | 76.0% | 90.3% |

| Table 2: Summary of career breaks for all respondents | | |
|---|---|---|
| | No career break | Yes career break |
| Female | 26.72% | 45.28% |
| Male | 73.28% | 54.72% |

| Table 3: Summary Statistics of Responses to the Three Survey Questions Regarding Promotions | | | | | | | | | | |
|---|---|---|---|---|---|---|---|---|---|---|
| | Female | | | | | Male | | | | |
| Question | Never | A few (0 - 5) | Many (5 -10) | A lot (10+) | More frequently than I recall | Never | A few (0 - 5) | Many (5 -10) | A lot (10+) | More frequently than I recall |
| How many times have you asked for a promotion (promotion means change in title)? | 35.82% | 61.94% | 0.74% | 1.49% | 0 | 31.51% | 61.48% | 5.84% | 1.17% | 0 |
| How many times have you received a promotion? | 17.16% | 76.12% | 6.72% | 0 | 0 | 9.73% | 75.49% | 12.84% | 1.56% | 0.39% |
| How many times have you been offered a promotion (without asking)? | 39.56% | 58.21% | 2.24% | 0 | 0 | 23.74% | 67.32% | 8.17% | 0.39% | 0.39% |

| Table 4: Summary Statistics of Responses to the Three Survey Questions Regarding Promotions – Collapsed Categories | | | | |
|---|---|---|---|---|
| | Female | | Male | |
| Question | Never | At least once | Never | At least once |
| How many times have you asked for a promotion (promotion means change in title)? | 36% | 64% | 32% | 68% |
| How many times have you received a promotion? | 17% | 83% | 10% | 90% |
| How many times have you been offered a promotion (without asking)? | 40% | 60% | 24% | 76% |

| Table 5: Summary of counts for all respondents to request a promotion (proportions of each gender identity cohort in brackets) | | | | | |
|---|---|---|---|---|---|
| | Never | A few(1-5) | Many(5-10) | A lot (10+) | All |
| Female | 48 (35.82%) | 83 (61.94%) | 1 (0.75%) | 2 (1.49%) | 134 (100%) |
| Male | 81 (31.40%) | 159 (61.63%) | 15 (5.81%) | 3 (1.16%) | 258 (100%) |
| Rather not specify | 4 (50.00%) | 2 (25.00%) | 2 (25.00%) | 0 (0.00%) | 8 (100%) |
| All | 133 (33.25%) | 244 (61.00%) | 18 (4.50%) | 5 (1.25%) | 400 (100%) |

| Table 6: Summary of proportions and expected proportions (in brackets) for all respondents with collapsed categories | | | |
|---|---|---|---|
| | Never | A few (1 - 5) | More than 5 |
| Female | 35.82% (33.28%) | 61.94% (60.97%) | 2.24% (5.75%) |
| Male | 31.39% (33.26%) | 61.3% (61.01%) | 6.98% (5.74%) |
| Rather not specify | 50.00% (33.75%) | 25% (61.25%) | 25.00% (6.25%) |

| Table 7: Proportions and expected proportions (in brackets) for all respondents in the portfolio management functional group, with 10-20 years of experience with a Masters/Doctorate with collapsed categories | | |
|---|---|---|
| | Never | At least once |
| Female | 20.00% (21.74%) | 80.00% (78.26%) |
| Male | 22.22% (21.74%) | 77.78% (78.26%) |

| Cohort | Everyone | | Respondents who did not take a career break | | Respondents who did take a career break | |
|---|---|---|---|---|---|---|
| Question and Outcome | Is promotion propensity independent of Gender? | Outcomes | Is promotion propensity independent of Gender? | Outcomes | Is promotion propensity independent of Gender? | Outcomes |
| **Propensity to request** (How many times have you asked for a promotion) | **Yes**. | Men and women ask for promotions in the same proportions | **Yes** | Men and women are asking for promotions in the same proportions. | **No** | Men who have taken a career break ask for promotions more than women who have taken a career break |
| **Propensity to be offered without asking** (How many times have you been offered a promotion without asking?) | **No** | Men receive more unsolicited promotions than women | **No** | Men receive more unsolicited promotions than women. | **No** | Men in this group are offered promotions without asking more than women in this group |
| **Propensity to receive a promotion when applied for** | **No** | Women receive more promotions when they request them | **No** | Women receive more promotions when they request them | **No** | Women receive more promotions when they request them |
| **Propensity to Receive** (How many times have you received a promotion regardless of whether it was offered or requested?) | **No** | Men receive more promotions than women overall regardless of whether the promotion was offered or requested. | **No** | Men receive more promotions than women | **No** | Men receive more promotions than women |

Table 8: Summary of the test results for the four promotion aspects